\documentclass[aps,prb,twocolumn,superscriptaddress,nofootinbib,longbibliography,floatfix]{revtex4-2}

\usepackage{graphicx,subfigure,bm,amssymb,amsmath,hyperref,dcolumn}
\usepackage{color,multirow,supertabular,float}
\usepackage{mathrsfs}
\usepackage{amsthm}
\usepackage{mathtools}
\usepackage{adjustbox}
\usepackage{tikz}

\begin{document}

\title{The $6-\epsilon$ Expansion for Long-Range Lee--Yang and Percolation Criticality}

\author{Zhiyi Li}
\affiliation{Department of Modern Physics, University of Science and Technology of China, Hefei, Anhui 230026, China}
\affiliation{Hefei National Laboratory, University of Science and Technology of China, Hefei 230088, China}

\author{Kun Chen}
        \email{chenkun@itp.ac.cn}
    \affiliation{CAS Key Laboratory of Theoretical Physics, Institute of Theoretical Physics, Chinese Academy of Sciences, Beijing 100190, China}

\author{Zhijie Fan}
\email{zfanac@ustc.edu.cn}
\affiliation{Hefei National Laboratory, University of Science and Technology of China, Hefei 230088, China}
\affiliation{Shanghai Research Center for Quantum Science and CAS Center for Excellence in Quantum Information and Quantum Physics, University of Science and Technology of China, Shanghai 201315, China}
\affiliation{Hefei National Research Center for Physical Sciences at the Microscale and School of Physical Sciences, University of Science and Technology of China, Hefei 230026, China}

\author{Youjin Deng}
\email{yjdeng@ustc.edu.cn}
\affiliation{Department of Modern Physics, University of Science and Technology of China, Hefei, Anhui 230026, China}
\affiliation{Hefei National Laboratory, University of Science and Technology of China, Hefei 230088, China}
\affiliation{Hefei National Research Center for Physical Sciences at the Microscale and School of Physical Sciences, University of Science and Technology of China, Hefei 230026, China}

\begin{abstract}
The crossover from long-range (LR) to short-range (SR) criticality in
percolation has remained unsettled because previous renormalization-group (RG) analysis within the $\epsilon'=3\sigma-d$ expansion
fixes the anomalous dimension at $\eta=2-\sigma$, whereas SR percolation
has $\eta_{\rm SR}<0$ near $d=6$.  Sak's matching condition then places the crossover above $\sigma=2$, outside the regime in which the LR
interaction dominates.  In spatial dimension $d=6-\epsilon$, we formulate a perturbative expansion for the LR $\phi^3$ field theory and perform a one-loop RG analysis throughout the perturbatively accessible nonclassical regime $0<\delta<\epsilon/3$, where $\delta = 2-\sigma$.  We derive the one-loop corrections to the critical exponents $\eta$ and $\nu$, which acquire nontrivial dependence on
$\epsilon$ and $\delta$. They reduce to their mean-field values at the LR upper critical line and continuously recover the SR $6-\epsilon$ results as $\sigma\to2$. These results support a crossover threshold
$\sigma_*=2$ and remove the apparent discontinuity of $\eta$
between the LR and SR values within this framework.  The same approach
also yields the anomalous and edge exponents of the LR Lee--Yang
universality class and $q$-state Potts universality classes with $q<2$.
\end{abstract}

\maketitle

\section{Introduction}
\label{sec:intro}

Long-range (LR) interactions are ubiquitous in nature, governing phenomena
ranging from dipolar ferromagnetism and screened Coulomb systems to the
dynamics of biophysical networks \cite{Defenu2023}.  They are commonly characterized by a power-law decay with distance,
$J(r)\sim r^{-(d+\sigma)}$, where $d$ is the spatial dimension and $\sigma$ controls the interaction range.
Recent advances have brought controllable LR interactions within 
reach in cold-atom, trapped-ion, and Rydberg-atom platforms 
\cite{Lahaye2009,Britton2012,Barredo2016,Bernien2017,Browaeys2020}. 
These interactions can give rise to phase transitions and critical 
behavior that differ markedly from those of short-range (SR) systems, 
forming $\sigma$-dependent universality classes. 
As $\sigma$ increases and the interaction range becomes effectively shorter, 
LR criticality eventually crosses over to SR 
universality at a threshold $\sigma_*$. 
The nature of LR critical behavior and the LR--SR crossover have 
drawn sustained attention, with the precise location of the 
boundary $\sigma_*$ standing out as the most debated issue for
over five decades.

Extensive research has focused on the LR $O(n)$-spin model, which consists of $n$-component spins on a lattice with such a power-law coupling and is described in the continuum limit by a $\phi^4$ theory with the fractional Laplacian $(-\Delta)^{\sigma/2}$ as the kinetic term.
Expanding this theory about its Gaussian fixed point in $\epsilon'=2\sigma-d$, Fisher, Ma, and Nickel found through a second-order RG analysis that the anomalous dimension $\eta$, defined via the critical two-point function $G(r)\sim r^{-(d-2+\eta)}$, remains exactly at its mean-field value,
\begin{equation}
    \eta = \eta_{\rm MF} = 2-\sigma,
\end{equation}
and suggested that this result might hold to all orders in $\epsilon'$ \cite{FisherMaNickel1972}.
By contrast, the SR $\epsilon=4-d$ expansion yields a nonzero $\eta_{\rm SR}$ already at two loops, so that naively extending the LR branch to $\sigma_*=2$ would produce a discontinuous jump in $\eta$.
Based on the assumption that $\eta= \eta_{\rm MF}$ strictly holds in the LR regime, and to avoid a jump in $\eta$, Sak proposed the LR--SR threshold $\sigma_*=2-\eta_{\rm SR}$, at which the LR and SR values of $\eta$ match continuously \cite{Sak1973}.
Monte Carlo studies of the two-dimensional LR Ising model reached opposite conclusions: some supported Sak's criterion after accounting for strong preasymptotic and finite-size corrections \cite{LuijtenBlote2002,AngeliniParisiRicciTersenghi2014,HoritaSuwaTodo2017}, while others favored $\sigma_*=2$ \cite{BlanchardPiccoRajabpour2013}. Most field-theoretic analyses have relied on expansions about $d=2\sigma$ or on conformal-field-theory approaches that assume $\eta=\eta_{\rm MF}$ without corrections \cite{HonkonenNalimov1989,Paulos2016,Behan2017,Benedetti2020,Benedetti2025}, though singular loop-integral behavior near $\sigma=2$ has occasionally been noted~\cite{Benedetti2020}. Recent systematic high-precision numerical studies of the Ising, XY, and Heisenberg models point to a common threshold $\sigma_*=2$~\cite{xiaoSpontaneousSymmetryBreaking2026,YaoXY2025,XiaoXY2025,LongRangeLERW2026,XiaoSak2026}. Meanwhile, we perform a complementary RG analysis within the $4-\epsilon$ expansion, which treats the entire non-mean-field regime as perturbatively accessible, in contrast to the $\epsilon'$ expansion near the mean-field line $d=2\sigma$. It provides two-loop corrections to $\eta$ at the Wilson--Fisher (WF) fixed point, which smoothly reduce to the SR perturbative result $\eta_{\rm SR}$, supporting the scenario $\sigma_*=2$ \cite{LiChenDeng2026}.

\begin{figure}[t]
\centering
\includegraphics[width=0.98\linewidth]{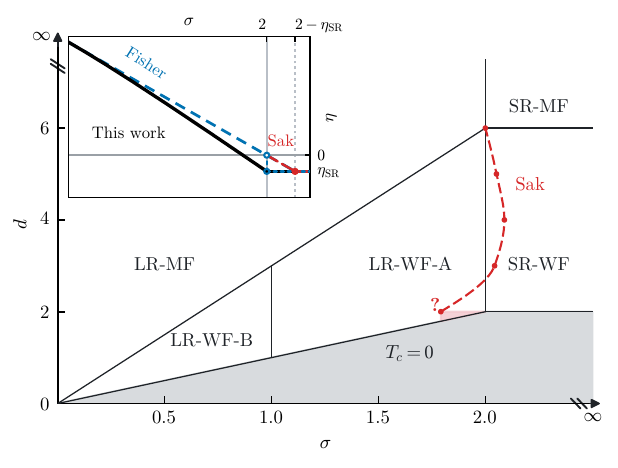}
\caption{\label{fig:percolation-crossover}%
Universality-class diagram for LR percolation in the $(\sigma,d)$ plane.
Each region denotes a universality class controlled by the indicated
renormalization-group fixed point: LR and SR mean long-range and
short-range, respectively; MF denotes mean field, and WF denotes the
Wilson--Fisher fixed point.  The two LR-WF sectors are separated at
$\sigma=1$: LR-WF-B is the $\sigma<1$ sector, where
$\eta=\eta_{\rm MF}$ is rigorously established, and LR-WF-A is the
$1<\sigma<2$ sector.  The boundaries $d=3\sigma$, $\sigma=2$, and $d=6$
separate LR-MF from LR-WF, LR from SR, and SR-WF from SR-MF,
respectively.
The gray region lies below the lower critical dimension and has
$T_c=0$.  In this
scenario the LR--SR boundary is $\sigma_*=2$ for all
$d$; the red dashed curve instead shows Sak's prediction
$\sigma_{\rm Sak}(d)=2-\eta_{\rm SR}(d)$.  The pale-red region marked
``?'' highlights the universality-class ambiguity left by Sak's
scenario for $43/24<\sigma<2$ and $\sigma<d<2$.  The inset compares the
anomalous dimension at $d=6-\epsilon$, with $\epsilon=0.2$ for illustration.  Fisher's
scenario (blue dashed) follows $\eta=\eta_{\rm MF}$ and jumps to
$\eta_{\rm SR}$ at $\sigma=2$; Sak's scenario (red dashed) follows
$\eta=\eta_{\rm MF}$ up to $\sigma=2-\eta_{\rm SR}$; and our result (black
solid) approaches $\eta_{\rm SR}$ continuously at $\sigma=2$.
}
\end{figure}


The LR percolation model provides another important playground for understanding critical phenomena with LR interactions. In LR bond percolation, two sites $i$ and $j$ on a $d$-dimensional hypercubic lattice are connected with a probability $p_{ij} \propto p/r_{ij}^{d+\sigma}$, where $r_{ij}$ is the distance between them. The connectivity transition is described by an $n$-component $\phi^3$ theory in the $n\to 0$ limit \cite{FortuinKasteleyn1972,PriestLubensky1976,Amit1976}.
Multi-loop RG calculations within the $6-\epsilon$ expansion give a negative SR anomalous dimension, $\eta_{\rm SR}<0$, and numerical simulations confirm this for all $3 \leq d<6$ \cite{PriestLubensky1976,Amit1976,Fisher1978}. For the LR case, an RG analysis using the $\epsilon'=3\sigma-d$ expansion about the upper critical dimension $d=3\sigma$ found no correction to $\eta=\eta_{\rm MF}$ through two loops \cite{TheumannGusmao1985}. If Sak's criterion $\sigma_*=2-\eta_{\rm SR}$ holds for percolation, the negative $\eta_{\rm SR}$ would give $\sigma_*>2$ [inset of Fig.~\ref{fig:percolation-crossover}].
This is unphysical because for $\sigma>2$ the SR kinetic term $k^2$ dominates over the LR term $|k|^\sigma$ by power counting, causing the LR field-theoretic RG to break down. To circumvent this difficulty, Theumann et al. proposed a sign-dependent prescription~\cite{TheumannGusmao1985}: Sak's criterion applies when $\eta_{\rm SR}>0$, while the crossover is placed at $\sigma_*=2$ with a discontinuity in the critical exponents when $\eta_{\rm SR} <0$. This prescription thus avoids an LR regime above $2$ only by reinstating the jump in $\eta$ that Sak's criterion was designed to eliminate---a circular resolution.

This controversy sharpens further in specific dimensions. In two dimensions, $\eta_{\rm SR}=5/24>0$, and Sak's criterion predicts $\sigma_*=43/24$ \cite{LiuXiaoFanDeng2025}. Earlier field theory for LR epidemic spreading, however, placed the boundary at $\sigma_*=2$ \cite{JanssenOerding1999}, and recent simulations of two-dimensional LR bond percolation have also found $\sigma_*=2$ \cite{LiuXiaoFanDeng2025}. Furthermore, it can be theoretically established that, in the supercritical phase ($p>p_c$), there exists a threshold $\sigma_*=2$ separating SR from LR epidemic-spreading behavior~\cite{grassberger2013two}. In $d=3,4,5$, SR percolation has $\eta_{\rm SR}<0$, so that Sak's boundary lies above $2$ and traces a nonmonotonic path in the $(\sigma,d)$ plane before returning to $2$ at the upper critical dimension $d=6$, as shown by the dashed line in Fig.~\ref{fig:percolation-crossover}~\cite{Borinsky2021}. Moreover, within the scenarios proposed by Sak and Theumann, it remains unclear which universality class should describe the regime $43/24<\sigma<2$ with $\sigma<d<2$, or equivalently, how the LR--SR crossover boundary connects to the boundary separating systems with and without a finite-temperature phase transition, as indicated by the red shaded region in Fig.~\ref{fig:percolation-crossover}.
Our recent study in Ref.~\cite{XiaoLiuFanDeng2026} proposed a unified $(\sigma,d)$-universality diagram for LR percolation by combining insights from Lévy flights~\cite{PhysRevLett.67.3047,JanssenOerding1999} with high-precision numerical simulations of LR $O(n)$ and percolation models~\cite{LiuXiaoFanDeng2025,xiaoSpontaneousSymmetryBreaking2026,YaoXY2025,XiaoXY2025,LongRangeLERW2026,XiaoSak2026}, as shown in Fig.~\ref{fig:percolation-crossover}. In this picture, the LR--SR boundary is set at $\sigma_*=2$ for all $d$, which naturally resolves the difficulties arising from Sak's criterion. Recent mathematical studies have reported that $\eta=\eta_{\rm MF}$ holds for $\sigma<1$~\cite{Hutchcroft2025CriticalI,Hutchcroft2025CriticalII,Hutchcroft2025CriticalIII,Hutchcroft2025Dimension}, and on this basis the diagram identifies $\sigma=1$ as an internal boundary within the LR non-mean-field regime.

Taken together, the LR--SR crossover raises two closely related questions: where the crossover boundary $\sigma_*$ lies and how the critical exponents, in particular the anomalous dimension $\eta_{\rm WF}(\sigma,d,n,m)$, behave across this boundary. Here, $m$ denotes the order of the leading interaction in the corresponding $\phi^m$ theory. Three distinct scenarios have been proposed to answer these questions: Fisher's scenario, Sak's scenario together with Theumann's modification, and our scenario. In Fisher's scenario, the boundary is fixed at $\sigma_*=2$, independent of $d$, $n$, and $m$, but the anomalous dimension sticks at the mean-field value, $\eta=\eta_{\rm MF}$ for $\sigma < 2$, leading to a jump in $\eta$ at $\sigma=2$. Sak's scenario avoids this jump by shifting the boundary to $\sigma_*=2-\eta_{\rm SR}$, thereby making the boundary dependent on the spatial dimension and universality class; for percolation, this results in a nonmonotonic boundary that extends into the unphysical regime $\sigma_*>2$ when $\eta_{\rm SR}<0$. Theumann's prescription modifies Sak's scenario to avoid this difficulty. In our scenario, the LR--SR boundary is fixed at $\sigma_*=2$, independent of $d$, $n$, and $m$, while $\eta_{\rm WF}(\sigma,d,n,m)$ within the LR regime retains a nontrivial dependence on these parameters, at least in the vicinity of $\sigma=2$. This leads to a simple and unified universality diagram. However, direct numerical evidence for the LR--SR crossover has so far been restricted mainly to $d=2$, which calls for a field-theoretic description, particularly within the $\phi^3$ theory of percolation.

Here we formulate a $d = 6-\epsilon$ expansion of the LR $\phi^3$ field theory. For small $\epsilon$, the entire non-mean-field regime is perturbatively accessible. Specifically, the non-mean-field LR regime $d/3<\sigma<2$ restricts $\delta=2-\sigma$ to $0<\delta<\epsilon/3$, so that $\delta$ remains perturbatively small throughout this window. Through a one-loop RG analysis, we obtain the WF fixed point and the anomalous dimension $\eta$ for LR percolation, as well as the correlation-length exponent $\nu$. The results are
\begin{subequations}
\label{critical exponent of per}
    \begin{align}
\eta_{\rm perc}
&=\delta-\frac{(\epsilon-3\delta)^2}
{3(7\epsilon-13\delta)}+O(\epsilon^2),
\label{eq:etaperc}\\
\frac{1}{\nu_{\rm perc}}
&=2-\delta-\frac{(\epsilon-3\delta)(5\epsilon-9\delta)}
{3(7\epsilon-13\delta)}+O(\epsilon^2),
\label{eq:nuperc}
\end{align}
\end{subequations}
which yield a nonzero correction relative to $\eta_{\rm MF}$ already at
the one-loop level as $\sigma\to2$.

Moreover, Equations~\eqref{critical exponent of per} match the conventional LR expansion as $\epsilon' =3\sigma-d = \epsilon-3\delta \to0$ and reduce continuously to the SR $6-\epsilon$ results as $\delta\to0$, as shown in the inset of Fig.~\ref{fig:percolation-crossover}. These results reveal that the apparent jump in $\eta$ that Sak's proposal aimed to eliminate does not arise in our framework. Therefore, the LR-WF fixed point persists for all $\sigma<2$ within the perturbative window, supporting the conclusion that the crossover occurs at $\sigma_*=2$, as shown in the universality diagram in Fig.~\ref{fig:percolation-crossover}. The same RG framework also determines the critical exponents of both the LR $q$-state Potts model with $q<2$ and the LR Lee--Yang edge singularity---the endpoint of partition-function zeros in the complex magnetic-field plane~\cite{YangLee1952,LeeYang1952,Fisher1978}.

\section{Models and Main Results}

\subsection{Models}
\label{sec:model}

We consider an $n$-component real scalar field
$\phi_a(x)$ governed by the effective action
\begin{align}
\mathcal{S}=\int d^dx\,\Big[&
\frac12\phi_a\big(r_0+K_L(-\Delta)^{\sigma/2}
+K_s(-\Delta)\big)\phi_a\nonumber\\
&+\frac{g_0}{3!}d_{abc}\phi_a\phi_b\phi_c\Big].
\label{eq:action}
\end{align}
Here, $a=1,\ldots,n$ labels the internal components, $x\in\mathbb{R}^d$,
and repeated internal indices are summed.  The spatial Laplacian is
denoted by $\Delta$; its fractional power
$(-\Delta)^{\sigma/2}$ has momentum-space kernel $|k|^\sigma$ and
represents pair couplings $J(r)\propto r^{-(d+\sigma)}$.  The
coefficients $K_L$ and $K_s$ multiply the LR kernel and the analytic
SR kernel $k^2$, respectively.  The bare mass $r_0$ tunes the transition,
with the critical point at $r_0=0$.  The parameter $g_0$ is the bare cubic
coupling, and the fully symmetric invariant tensor $d_{abc}$ contracts
the three fields into a scalar under the internal symmetry.

The tensor contractions define the symmetry coefficients $A$ and $B$
by $d_{acd}d_{bcd}=A\delta_{ab}$ and
$d_{ade}d_{bef}d_{cfd}=B d_{abc}$, respectively, where
$\delta_{ab}$ is the Kronecker delta.  We use their
normalization-independent ratio $\rho=B/A$ below.

\emph{Lee--Yang universality class.}
Above the critical temperature of a ferromagnet, partition-function
zeros on the imaginary magnetic-field axis terminate at the Yang--Lee
edge \cite{YangLee1952,LeeYang1952,KortmanGriffiths1971}.  The edge
singularity is described by the scalar field of
Eq.~\eqref{eq:action} with a purely imaginary cubic coupling such that $g_0^2<0$
\cite{Fisher1978}; setting $d_{111}=1$ then gives $\rho_{\rm LY}=1$.

\emph{Percolation and the $q$-state Potts universality class.}
Independent
percolation is the $q\to1$ limit of the $q$-state Potts model in the
Fortuin--Kasteleyn representation \cite{FortuinKasteleyn1972,Potts1952}.
Its continuum order parameter is described by an $n$-component field,
where $n=q-1$.

To specify the cubic invariant, let
$\bm e^\alpha=(e_1^\alpha,\ldots,e_{q-1}^\alpha)$ denote the Potts
vector assigned to state $\alpha=1,\ldots,q$; thus $e_a^\alpha$ is
its component along field direction $a$.  The $q$ vectors point to the
vertices of a regular hypertetrahedron in $\mathbb{R}^{q-1}$ and obey
$\sum_\alpha e_a^\alpha=0$,
$\bm e^\alpha\!\cdot\bm e^\beta=q\delta^{\alpha\beta}-1$, and
$\sum_\alpha e_a^\alpha e_b^\alpha=q\delta_{ab}$, where Greek
indices label Potts states, whereas Latin indices label field
components. 
These identities yield the group-theoretical factors
$A(q)=q^2(q-2)$ and $B(q)=q^2(q-3)$. From the definition of $A(q)$,
one obtains $d_{abc}d_{abc}=(q-1)A(q)$,
which vanishes at both $q=1$ and $q=2$, albeit for different reasons. At $q=1$, the vanishing originates from the formal number of field components, $n=q-1=0$. Percolation is therefore described by the analytic
continuation of the Potts theory to the limit $q\to1$. By contrast,
at $q=2$, one has $d_{abc}=0$, so that the cubic interaction vanishes
identically. The Potts model then reduces to the Ising model, whose SR critical behavior is described by $\phi^4$ theory with upper critical dimension $d_c=4$. Nevertheless, recent work has shown that the FK-Ising model exhibits an additional geometric upper critical dimension $d_u=6$~\cite{Fang_2022}. More generally, for $q<2$, the
FK-Potts model is expected to have the geometric upper critical
dimension $d_u=6$, with the ratio
\begin{equation}
    \rho=\frac{3-q}{2-q}.
    \label{eq:rho}
\end{equation}
The percolation limit $q\to1$ then gives $\rho=2$.

Let $t$ be the relevant distance from criticality. For percolation and
the Potts model, the constant $p$ controls the bond-occupation probability,
and $t\propto p_c-p$; for the LY edge,
$t\propto h_c-h_0$, with $h=ih_0$ the imaginary magnetic field and
$ih_c$ its edge value.  In the field theory,
$t\propto r_0-r_{0c}$, where $r_{0c}$ is the critical bare mass.  The
correlation length $\xi$ diverges as $\xi\sim|t|^{-\nu}$, which
defines the critical exponent $\nu$.  At criticality, the momentum-space two-point function
$G_{ab}(k)=\langle\phi_a(k)\phi_b(-k)\rangle
=\delta_{ab}G(k)$ scales as $G(k)\propto|k|^{-(2-\eta)}$ for
$k\to0$, defining the anomalous dimension $\eta$.  The LR Gaussian
propagator $G(k)\propto|k|^{-\sigma}$ therefore has
$\eta=\eta_{\rm MF}$.

For the LY universality class, the density of zeros $\varrho(h_0)$ behaves as
$\varrho(h_0)\sim(h_0-h_c)^{\sigma_{\rm edge}}$ as the edge is
approached from above, defining the edge exponent
$\sigma_{\rm edge}$.  With field scaling dimension
$\Delta_\phi=(d-2+\eta)/2$, scale invariance gives
\begin{equation}
\sigma_{\rm edge}=\frac{d-2+\eta_{\rm LY}}{d+2-\eta_{\rm LY}} \label{edgeexponent}.
\end{equation}

The action, tensor algebra, and observables now specify the two
universality classes.  We next identify the perturbative regime by
tree-level power counting and then present the one-loop RG results.

\subsection{Main results}
\label{sec:results}

A tree-level dimensional analysis of Eq.~\eqref{eq:action}, with
momentum dimension $[p]=1$ and $[K_L]=0$, gives
$[\phi]=(d-\sigma)/2$, $y_t=\sigma$, $y_g=(3\sigma-d)/2$, and
$y_{K_s}=\sigma-2$ at the LR Gaussian fixed point.  Hence the LR upper
critical dimension is $d_{\rm up}^{\rm LR}=3\sigma$, while the SR
kinetic term is irrelevant for $\sigma<2$.

Writing $d=6-\epsilon$ and $\sigma=2-\delta$, the canonical dimension
of the cubic coupling at this fixed point is
$y_g=\epsilon'/2$, where
$\epsilon'\equiv3\sigma-d=\epsilon-3\delta$.  The interacting LR regime
$d<3\sigma$ together with $\sigma<2$ therefore corresponds to
$0<\delta<\epsilon/3$, which is perturbatively accessible for
$\delta=O(\epsilon)$.  We perform the one-loop RG analysis using the
dimensionless renormalized coupling $u=g^2/(4\pi)^{d/2}$.

Solving the one-loop beta function yields a nontrivial zero $u^*$,
the LR Wilson--Fisher fixed point (LR-WFP).  In terms of the
normalization-independent combination $Au$, the fixed point and its RG
eigenvalue $y_u$ along the coupling direction are
\begin{subequations}
    \label{fixpoint}
\begin{align}
Au^*&=-\frac{2(\epsilon-2\delta)\epsilon'} 
{4\rho(\epsilon-2\delta)-\epsilon'}+O(\epsilon^2), \label{eq:ustar}\\
 y_u&=-\epsilon'+O(\epsilon^2).
\end{align}
\end{subequations}
Since $\epsilon'=\epsilon-3\delta>0$ throughout $0<\delta<\epsilon/3$, $y_u<0$, and
the LR-WFP is stable in the infrared within the critical surface.
Moreover, the Lee--Yang, percolation, and Potts ($q<2$)
fixed points satisfy $Au^*<0$, providing a common signature of their nonunitary character.
In the Lee--Yang theory, this sign arises from an imaginary coupling, whereas in percolation the coupling remains real and $A<0$, indicating an indefinite field-space metric and the presence of negative-norm states. 

For the general $\phi^3$ theory, the critical exponents depend on the internal symmetry through the normalization ratio $\rho$ as follows:
\begin{subequations}
    \label{criticalexponent_main}
\begin{align}
\eta&=\delta-
\frac{(\epsilon-3\delta)^2}
{3\left[4\rho(\epsilon-2\delta)-(\epsilon-3\delta)\right]}
+O(\epsilon^2),
\label{eq:etamain}\\
\frac{1}{\nu}&=\sigma-
\frac{(\epsilon-3\delta)(5\epsilon-9\delta)}
{3\left[4\rho(\epsilon-2\delta)-(\epsilon-3\delta)\right]}
+O(\epsilon^2).
\label{eq:numain}
\end{align}
\end{subequations}
These results apply to the $q$-state FK-Potts model, with $\rho(q)$ given by Eq.~\eqref{eq:rho}. For independent percolation ($\rho=2$), they reduce to Eqs.~\eqref{critical exponent of per}. For the Lee--Yang universality class ($\rho=1$), they give
\begin{equation}
   \eta_{\rm LY}
=\delta-\frac{(\epsilon-3\delta)^2}
{3(3\epsilon-5\delta)}+O(\epsilon^2),
\label{eq:etaLY}
\end{equation}
The corresponding edge exponent follows by substituting Eq.~\eqref{eq:etaLY} into the scaling relation in Eq.~\eqref{edgeexponent}.
The mass eigenvalue
$1/\nu_{\rm LY}$ contains a redundant contribution associated with a constant field shift, so the physical LY critical behavior is more directly characterized by $\eta_{\rm LY}$ and the edge exponent $\sigma_{\rm edge}$.  


At the SR boundary $\delta\to0$, Eqs.~\eqref{eq:etamain}
and~\eqref{eq:numain} give
$\eta_{\rm SR}=-\epsilon/[3(4\rho-1)]$ and
$1/\nu_{\rm SR}=2-5\epsilon/[3(4\rho-1)]$.
The choices $\rho=1$ and
$2$ reproduce, respectively, the standard SR results
$\eta_{\rm LY}=-\epsilon/9$ and
$\eta_{\rm perc}=-\epsilon/21$, together with the corresponding
correlation-length exponents $1/\nu_{\rm LY}=2-5\epsilon/9$ and $1/\nu_{\rm perc}=2-5\epsilon/21$
\cite{PriestLubensky1976,Amit1976,Fisher1978,BonfimKirkhamMcKane1981,Gracey2015}.

Near the LR mean-field boundary $\epsilon'\to0$ with $\delta$ fixed, the
expansion gives
$\eta-\delta=-\epsilon'^2/(12\rho\delta)+O(\epsilon'^3)$ and
$1/\nu=\sigma-\epsilon'/(2\rho)+O(\epsilon'^2)$.  Thus $\eta$
approaches $\eta_{\rm MF}$ quadratically in $\epsilon'$, while $1/\nu$
agrees with the conventional fixed-$\sigma$ expansion about
$d=3\sigma$ \cite{TheumannGusmao1985}.  At $\epsilon'=0$, the critical
exponents take their LR mean-field values,
$\eta=\eta_{\rm MF}=\delta$ and $1/\nu=\sigma$.

Notably, the denominator in
Eqs.~\eqref{eq:etamain} and~\eqref{eq:numain} vanishes at
$\delta/\epsilon=3/5$ for LY and $7/13$ for percolation, both outside the LR-WF regime
$0<\delta<\epsilon/3$.  The fixed point and its exponents are
therefore regular throughout the perturbative LR window.

Equation~\eqref{eq:etamain} gives
$\eta-\eta_{\rm MF}<0$ for $0<\delta<\epsilon/3$.  As the LR window is
traversed from $d=3\sigma$ to $\sigma=2$, $\eta$ evolves from its LR
mean-field value to the negative SR value, rather than remaining
locked at $\eta_{\rm MF}$, as shown in Fig.~\ref{fig:percolation-crossover}.  Therefore, for the universality classes discussed here, our results support the scenario with the LR--SR boundary $\sigma_*=2$.



Having stated the fixed point, its limiting checks, and the crossover
inference, we now present our systematic RG analysis with counterterms and RG functions that lead to
these results.

\begin{figure}[b]
\centering
\begin{tikzpicture}[line width=0.8pt,scale=0.82]
\begin{scope}[shift={(0,0)}]
  \draw (-1.45,0) -- (-0.58,0);
  \draw (0.58,0) -- (1.45,0);
  \draw (0,0) circle (0.58);
  \filldraw (-0.58,0) circle (1.7pt);
  \filldraw (0.58,0) circle (1.7pt);
  \node at (-1.2,0.28) {$k$};
  \node at (0,0.88) {$p$};
  \node at (0,-0.92) {$p+k$};
  \node at (0,-1.6) {(a)};
\end{scope}
\begin{scope}[shift={(3.85,-0.1)}]
  \coordinate (v1) at (90:0.62);
  \coordinate (v2) at (210:0.62);
  \coordinate (v3) at (-30:0.62);
  \draw (v1) -- (v2) -- (v3) -- cycle;
  \draw (v1) -- (90:1.35);
  \draw (v2) -- (210:1.35);
  \draw (v3) -- (-30:1.35);
  \filldraw (v1) circle (1.7pt);
  \filldraw (v2) circle (1.7pt);
  \filldraw (v3) circle (1.7pt);
  \node at (0.42,1.1) {$k_1$};
  \node at (-1.5,-0.75) {$k_2$};
  \node at (1.52,-0.75) {$k_3$};
  \node at (0,-1.5) {(b)};
\end{scope}
\begin{scope}[shift={(7.4,0)}]
  \draw (-1.45,0) -- (-0.58,0);
  \draw (0.58,0) -- (1.45,0);
  \draw (0,0) circle (0.58);
  \filldraw (-0.58,0) circle (1.7pt);
  \filldraw (0.58,0) circle (1.7pt);
  \draw[fill=white] (0,0.58) circle (0.14);
  \draw (-0.099,0.481) -- (0.099,0.679);
  \draw (-0.099,0.679) -- (0.099,0.481);
  \node at (0,-1.6) {(c)};
\end{scope}
\end{tikzpicture}
\caption{One-loop diagrams of the LR cubic theory:
(a) the self-energy bubble, (b) the vertex triangle, and
(c) the bubble with one mass insertion (crossed circle).
Solid lines denote the LR propagator
$G_0(k)=(\mu^\sigma r+|k|^\sigma)^{-1}$, and dots denote the vertex
$-g\mu^{\epsilon'/2}d_{abc}$.}
\label{fig:diagrams}
\end{figure}
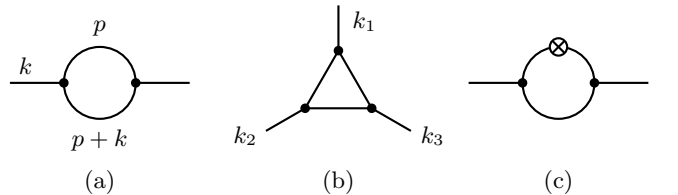

\section{Method}
\label{sec:method}
\subsection{Renormalization scheme}
\label{sec:scheme}
For $\sigma<2$, we set the irrelevant SR amplitude $K_s=0$ and
normalize $K_L=1$.  At the renormalization momentum scale $\mu$, the action is
\begin{align}
\mathcal{S}={}&\int d^dx\,\Big[
\frac12\phi_a\big(\mu^\sigma r+(-\Delta)^{\sigma/2}\big)\phi_a
\nonumber\\
&\hspace{24mm}
+\frac{g\mu^{\epsilon'/2}}{3!}d_{abc}\phi_a\phi_b\phi_c
\Big]\nonumber\\
&+\int d^dx\,\Big[
\frac12\phi_a\big(\mu^\sigma\delta r
+\delta Z(-\Delta)^{\sigma/2}\big)\phi_a
\nonumber\\
&\hspace{24mm}
+\frac{\delta g\mu^{\epsilon'/2}}{3!}d_{abc}\phi_a\phi_b\phi_c
+\delta h_a\phi_a\Big].
\label{eq:renaction}
\end{align}
Here $r$ is the dimensionless renormalized mass; $\delta r$, $\delta Z$, and $\delta g$ are the
counterterms for the mass, field, and interaction, respectively.  The source counterterm $\delta h_a$ enforces
$\langle\phi_a\rangle=0$ and removes tadpoles from one-particle
irreducible (1PI) vertex functions.  Defining
$Z_\phi=1+\delta Z$, the bare quantities satisfy
$\phi_{0a}=Z_\phi^{1/2}\phi_a$,
$g_0=\mu^{\epsilon'/2}(g+\delta g)Z_\phi^{-3/2}$, and
$r_0=\mu^\sigma(r+\delta r)Z_\phi^{-1}$. For simplicity, we use $\epsilon'=\epsilon-3\delta$ below, although the expansion itself is performed in $\epsilon=6-d$ rather than $\epsilon'$.

Let $\Gamma_2(k)$ denote the scalar part of the 1PI two-point vertex,
$\Gamma^{(2)}_{ab}(k)=\delta_{ab}\Gamma_2(k)$, and let $\Gamma_3$
denote the scalar coefficient of the 1PI three-point vertex,
$\Gamma^{(3)}_{abc}=d_{abc}\Gamma_3$.  We denote the 1PI self-energy
by $\Sigma(k;r)$, so that
$\Gamma_2(k)=\mu^\sigma(r+\delta r)+Z_\phi|k|^\sigma-\Sigma(k;r)$.
We impose the renormalization conditions
\begin{subequations}
    \begin{align}
\Gamma_2(0)&=\mu^\sigma r \label{rcondition},\\
\left.\frac{\partial\Gamma_2}{\partial|k|^\sigma}
\right|_{|k|=\mu}&=1 \label{etacondition},\\
\left.\Gamma_3(k_1,k_2,k_3)\right|_{\rm sym}
& =-g\mu^{\epsilon'/2}\label{gcondition},
\end{align}
\label{eq:renconditions}
\end{subequations}
where the symmetric point satisfies
$k_1+k_2+k_3=0$ and $|k_1|=|k_2|=|k_3|=\mu$.
These conditions determine $\delta r$, $\delta Z$, and $\delta g$ by calculating the two-point and three-point vertex diagrams order by order.

\subsection{One-loop integrals and counterterms}
\label{sec:oneloop}

Specifically, each diagram factorizes into a tensor contraction and a scalar momentum
integral. At the one-loop level, the bubble diagram (Fig.~\ref{fig:diagrams}(a)) carries $A$, while the triangle (Fig.~\ref{fig:diagrams}(b)) carries $B$.  For simplicity, we first consider the massless case at $r=0$, where the system is critical, using the following massless convolution identity, in which $\Gamma$
denotes the Euler gamma function:
\begin{align}
&\int\frac{d^dp}{(2\pi)^d}
\frac{1}{|p|^{2a}|p+k|^{2b}}
=\frac{|k|^{d-2a-2b}}{(4\pi)^{d/2}}\nonumber\\
&\quad\times
\frac{\Gamma(\tfrac d2-a)\Gamma(\tfrac d2-b)
\Gamma(a+b-\tfrac d2)}
{\Gamma(a)\Gamma(b)\Gamma(d-a-b)}.
\label{eq:master}
\end{align}
Here $p$ is the loop momentum, $k$ the external momentum, and $a$ and
$b$ are the propagator exponents.

\emph{Self-energy bubble.}
At criticality, the one-loop self-energy
$\Sigma^{(1)}(k)$ from Fig.~\ref{fig:diagrams}(a) is
\begin{align}
\Sigma^{(1)}(k)
&=\frac{Ag^2\mu^{\epsilon'}}{2}
\int\frac{d^dp}{(2\pi)^d}
\frac{1}{|p|^\sigma|p+k|^\sigma}\nonumber\\
&\simeq-\frac{Au}{6(\epsilon-2\delta)}
|k|^\sigma\left(\frac{|k|}{\mu}\right)^{-\epsilon'}.
\label{eq:sigma1}
\end{align}
The last line follows from Eq.~\eqref{eq:master} and
$\Gamma[-1+(\epsilon-2\delta)/2]
=-2/(\epsilon-2\delta)+O(1)$.  The pole appears because $\delta$ is a small parameter of order $\epsilon$ in our expansion, which means that the loop power $|k|^{d-2\sigma}$ approaches the analytic kernel $k^2$.  As a result, this pole unlocks $\eta$ from $\eta_{\rm MF}$.
The renormalization condition in Eq.~\eqref{etacondition} gives
$Z_\phi=1-z_1u+O(u^2)$ with
$z_1=A/[6(\epsilon-2\delta)]$.

\emph{Vertex triangle.}
Let $T_3$ denote the scalar triangle integral in
Fig.~\ref{fig:diagrams}(b).  At the symmetric point, its ultraviolet
pole is
\begin{align}
T_3&=\int\frac{d^dp}{(2\pi)^d}
\frac{1}{|p|^\sigma|p+k_1|^\sigma}\nonumber
\frac{1}{|p+k_1+k_2|^\sigma}\bigg|_{\rm sym}\nonumber\\
&=\frac{\mu^{-\epsilon'}}{(4\pi)^3}
\left[\frac{1}{\epsilon'}+O(1)\right].
\label{eq:T3}
\end{align}
The vertex renormalization
condition in Eq.~\eqref{gcondition} then yields
\begin{equation}
    g_0=\mu^{\epsilon'/2}gZ_\phi^{-3/2}
\left[1-\frac{Bu}{\epsilon'}+O(u^2)\right].
\label{eq:g0}
\end{equation}

\emph{Mass insertion.}
The mass insertion in Fig.~\ref{fig:diagrams}(c) is a diagrammatic
representation of the first-order mass expansion of the self-energy.
The crossed circle denotes the mass $r$. The diagram gives
\begin{equation}
    \frac{\partial\Sigma}{\partial r}\bigg|_{r=0}\supset
-A\,g_0^2\int_{|p|>\mu}\frac{d^dp}{(2\pi)^d}\frac{1}{|p|^{3\sigma}}
=-\frac{A\,u}{\epsilon'}.
\end{equation}
The mass renormalization condition in Eq.~\eqref{rcondition} then yields
\begin{equation}
    r_0=\mu^\sigma r
\left[1-\frac{Au}{\epsilon'}+z_1u+O(u^2)\right].
\label{eq:r0}
\end{equation}

\subsection{RG functions and exponents}
\label{sec:rgfunctions}

Define the bare loop coupling
$u_0=g_0^2/(4\pi)^{d/2}$.  Requiring $u_0$ and $r_0$ to be independent
of $\mu$ defines
$\beta_u=\mu\,\partial_\mu u|_{g_0,r_0}$ and
$\beta_r=\mu\,\partial_\mu r|_{g_0,r_0}$.  To one loop,
\begin{align}
\beta_u&=-\epsilon'u+
\left[\frac{A\epsilon'}{2(\epsilon-2\delta)}-2B\right]u^2
+O(u^3),
\label{eq:betau}\\
\beta_r&=-\sigma r-Aur
\left[1-\frac{\epsilon'}{6(\epsilon-2\delta)}\right]
+O(u^2r).
\label{eq:betar}
\end{align}
The field-renormalization contribution to $\beta_u$ is of the same
order as the vertex contribution and must be retained.  The nontrivial
zero of Eq.~\eqref{eq:betau} is the LR-WFP given by Eq.~\eqref{eq:ustar}.

Define the field anomalous-dimension function by
$\gamma=\tfrac12\mu\,\partial_\mu\ln Z_\phi$, with the bare parameters
held fixed.  At one loop,
$\gamma(u)=A\epsilon'u/[12(\epsilon-2\delta)]+O(u^2)$.  The
Callan--Symanzik equation for the two-point function then gives
$\eta=\eta_{\rm MF}+2\gamma(u^*)$.  Similarly, linearizing
Eq.~\eqref{eq:betar} gives
$1/\nu=-\partial_r\beta_r|_{u^*}$.  Substitution of the fixed point
reproduces Eqs.~\eqref{eq:etamain} and~\eqref{eq:numain}.

\section{Discussion}
\label{sec:discussion}

We have developed a $d=6-\epsilon$ expansion for LR $\phi^3$ theories
and obtained the infrared-stable LR fixed point and the one-loop
exponents for the percolation, Lee--Yang edge, and $q<2$ FK-Potts
universality classes.  The results interpolate between the LR
mean-field line $d=3\sigma$ and the SR $6-\epsilon$ expansion at
$\sigma=2$.  They support the scenario summarized in
Fig.~\ref{fig:percolation-crossover}: the critical exponents may depend
on $(d,\sigma,n,m)$, whereas the crossover location $\sigma_*=2$ is
independent of $d$, $n$, and the interaction order $m$.

A recent mathematical study provides a complementary perspective on
this problem. For $0<\sigma<d$, Ref.~\cite{Hutchcroft2022Sharp} reports
an averaged upper bound on the critical two-point function obtained using
a hierarchical decomposition, multiscale renormalization, and percolation
inequalities.  The reported bound is argued to imply $\eta\geq\eta_{\rm MF}$
and to support the locked branch assumed
in Sak's scenario, whereas our one-loop RG calculation gives
$\eta<\eta_{\rm MF}$.  The distinction becomes especially consequential
for percolation in $3\leq d<6$, where $\eta_{\rm SR}<0$ would place
Sak's threshold $2-\eta_{\rm SR}$ above $2$.  From the RG-stability
perspective, Theumann and Gusm\~ao instead found the LR fixed point to
be stable for all $\sigma<2$ and the LR expansion to become ill-defined
beyond $\sigma=2$, where the analytic $k^2$ term dominates
\cite{TheumannGusmao1985}.  How these mathematical and RG descriptions
can be reconciled therefore remains an open question, calling for a
rigorous analysis of the crossover and further independent checks of
the perturbative calculations in this work as well as the mathematical analysis in Ref.~\cite{Hutchcroft2022Sharp}.

Our expansion is controlled only near six dimensions and, in low dimensions, can only provide a quantitative, rather than rigorous, picture.  It also does not give a controlled fixed-$d$ description
near $d=3\sigma$ away from the vicinity of $(\sigma,d)=(2,6)$.  Future
work should extend the calculation to higher loops, determine the
physics at $\sigma_*=2$---including possible logarithmic
corrections---and clarify whether the LR regime $\sigma<2$ admits a CFT
description.

\begin{acknowledgments}
The authors thank Ziyu Liu for valuable discussions.
K.C. is supported by the National Key Research and Development Program of China, Grant No. 2024YFA1408604, the National Natural Science Foundation of China under Grant Nos. 12474245 and 12447103, and the GHfund A (202407010637).
Z.L., Z.F., and Y.D. are supported by the National Natural Science Foundation of China under Grant No. 12275263 and the Quantum Science and Technology-National Science and Technology Major Project under Grant No. 2021ZD0301900. Y.D. is also supported by the Natural Science Foundation of Fujian Province of China under Grant No. 2023J02032. Z.F. is also supported by the NSFC under Grant No. 12504265.
\end{acknowledgments}


\bibliography{references}

\end{document}